\documentclass{article}

\usepackage{arxiv}

\usepackage[utf8]{inputenc} 
\usepackage[T1]{fontenc}    
\usepackage{hyperref}       
\usepackage{url}            
\usepackage{booktabs}       
\usepackage{amsfonts}       
\usepackage{nicefrac}       
\usepackage{microtype}      
\usepackage{lipsum}		
\usepackage{graphicx}
\usepackage{natbib}
\usepackage{doi}
\usepackage{xcolor}

\title{A Stock Trading System for a Medium Volatile Asset using Multi Layer Perceptron}

\author{ \href{https://orcid.org/0000-0002-3843-386X}{\includegraphics[scale=0.06]{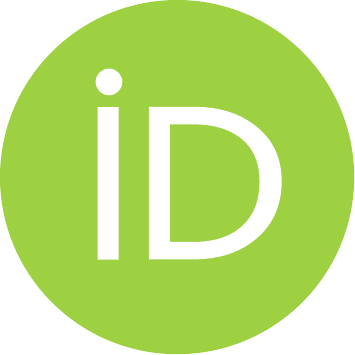}\hspace{1mm}Ivan Letteri}\thanks{\url{https://www.ivanletteri.it} } \\
	Department of Engineering, Computer Science and Maths\\
	University of L'Aquila\\
	L'Aquila, Italy, 67100 \\
	\texttt{ivan.letteri@univaq.it} \\
	\And
	\href{https://orcid.org/0000-0003-2327-9393}{\includegraphics[scale=0.06]{orcid.pdf}\hspace{1mm}Giuseppe Della Penna} \\
	Department of Engineering, Computer Science and Maths\\
	University of L'Aquila\\
	L'Aquila, Italy, 67100 \\
	\texttt{giuseppe.dellapenna@univaq.it} \\
	\And
    \href{https://orcid.org/0000-0001-9521-4711}{\includegraphics[scale=0.06]{orcid.pdf}\hspace{1mm}Giovanni De Gasperis}\\
	Department of Engineering, Computer Science and Maths\\
	University of L'Aquila\\
	L'Aquila, Italy, 67100 \\
	\texttt{giovanni.degasperis@univaq.it} \\
	\And
    \href{https://orcid.org/0000-0003-0329-2419}{\includegraphics[scale=0.06]{orcid.pdf}\hspace{1mm}Abeer Dyoub}\\
	Department of Engineering, Computer Science and Maths\\
	University of L'Aquila\\
	L'Aquila, Italy, 67100 \\
	\texttt{abeer.dyoub@univaq.it} \\
}

\date{}

\renewcommand{\undertitle}{}
\renewcommand{\shorttitle}{\textit{arXiv} Template}

\hypersetup{
pdftitle={A template for the arxiv style},
pdfsubject={ai-algo trading.NC, ai-algo trading.QM},
pdfauthor={Ivan Letteri et al.},
pdfkeywords={First keyword, Second keyword, More},
}

\begin{document}
\maketitle

\begin{abstract}
Stock market forecasting is a lucrative field of interest with promising profits but not without its difficulties and for some people could be even causes of failure. Financial markets by their nature are complex, non-linear and chaotic, which implies that accurately predicting the prices of assets that are part of it becomes very complicated. In this paper we propose a stock trading system having as main core the feed-forward deep neural networks (DNN) to predict the price for the next 30 days of open market, of the shares issued by Abercrombie \& Fitch Co. (ANF) in the stock market of the New York Stock Exchange (NYSE).

The system we have elaborated calculates the most effective technical indicator, applying it to the predictions computed by the DNNs, for generating trades. The results showed an increase in values such as Expectancy Ratio of 2.112\% of profitable trades with Sharpe, Sortino, and Calmar Ratios of 2.194, 3.340, and 12.403 respectively. As a verification, we adopted a backtracking simulation module in our system, which maps trades to actual test data consisting of the last 30 days of open market on the ANF asset. Overall, the results were promising bringing a total profit factor of 3.2\% in just one month from a very modest budget of \$100. This was possible because the system reduced the number of trades by choosing the most effective and efficient trades, saving on commissions and slippage costs.
\end{abstract}


\section{Introduction}
Stock market forecasting is considered a research field with promising returns for investors. However, there are considerable challenges to predicting stock market trends accurately and precisely enough due to their chaotic and non-linear nature, and also complexity. At the state of the art, there are numerous approaches for generating, processing, and optimizing a dataset, such as (\cite{LetteriPVG20}). 

Many artificial intelligence methods have been employed to classify cyber attacks (\cite{LetteriPG18}), predict network traffic anomalies (\cite{LetteriPG19}), course of a disease, and even the price trend in the stock market. Artificial neural networks (ANNs) to date remain a fairly popular choice for these kinds of tasks and are widely studied (\cite{LetteriPC19}), having exhibited good performance (\cite{Yetis2014StockMP}).

In particular, although deep neural networks (DNNs) are predominantly used for image recognition and natural language processing, showing surprising results (\cite{Soniya2015ARO}), they have also been applied to financial markets for stock price prediction using textual news analysis (\cite{Day2016DeepLF}).

In this paper, we propose a trading system for the stock market that uses forecasts by four Price Action (PA)-based DNNs to generate trading signals, and subsequently evaluate their performance in the stock market. This constitutes part of a work that aims to convert the DNNs forecasts into a profitable investment and trading system that can be automated, using a robot advisor.

\section{Data Set}
\label{sec:methodology}
This section describes the methodology with which the dataset is collected, the criteria adopted for the statistical analysis of the data collected, the training process and the series of tests conducted on the DNNs with optimized processes (\cite{LetteriDSopt2020}). In the next section we will show the trading rules applied on the forecast output with the related tests on profitability verified by our backtracking system.

\subsection{Technical Analysis}
\label{sec:dataset}
\begin{figure}[!ht]
	\centerline{\includegraphics[width=40em]{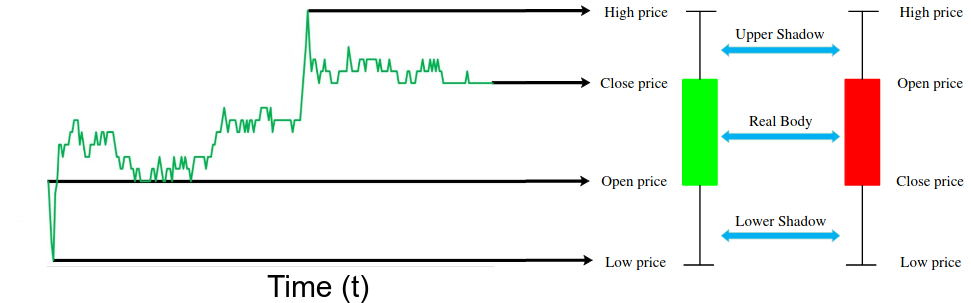}}
	\caption{Example of candlestick chart.}
	\label{fig:candlestick}
\end{figure}

Technical analysis (TA) constitutes the type of investment analysis that uses simple mathematical formulations or graphical representations of the time series of financial assets to explore trading opportunities. In its algorithmic form, TA uses the analysis of asset price history series (\cite{wang2021forecasting}), defined as OHLC, i.e., the opening, highers, lowest and closing prices of an asset, typically represented with candlesticks charts (see, e.g. in fig. \ref{fig:candlestick}). For each timeframe $t$, the OHLC of an asset is represented as a 4-dimensional vector
$X_t = (x^{(o)}_t,x^{(h)}_t,x^{(l)}_t,x^{(c)}_t)^T$, where $x_t^{(l)} > 0$, $x_t^{(l)} < x_t^{(h)}$ and  $x_t^{(o)}, x_t^{(c)} \in [x_t^{(l)},x_t^{(h)}]$.

Some stocks are more volatile than others. For example the shares of a large blue-chip company may not have large price fluctuations and are therefore said to have \textit{low volatility}, whereas the shares of a tech stock may fluctuate often and therefore have \textit{high volatility}. There are also stocks with \textit{medium volatility}\footnote{https://www.fool.com/investing/how-to-invest/stocks/stock-market-volatility/}, and this is the case of the asset that we studied in this article.

The dataset used in this work consists of historical OHLC prices data from the New York Stock Exchange (NYSE), the world's largest stock exchange by trading volume and the second largest by number of listed companies. Its share volume was surpassed by NASDAQ in the 1990s, but the total capitalization of the 2800 companies on the NYSE is five times that of the competing technology exchange. The asset on which we conducted our study is the shares issued by Abercrombie \& Fitch Co. (ANF). From a summary of results for the third quarter ended October 30, 2021 compared to the third quarter ended October 31, 2020, ANF reported net sales of \$905 million, up 10\% year-over-year and 5\% compared to net sales in the third quarter pre-COVID 2019. Digital net sales of \$413 million increased 8\% compared to last year and 55\% compared to pre-COVID 2019 third quarter digital net sales. Gross profit rate of 63.7\%, down approximately 30 basis points from last year and up approximately 360 basis points from 2019 (source GlobeNewswire\footnote{www.globenewswire.com/news-release/2021/11/23/2339734/0/en/Abercrombie-Fitch-Co-Reports-Third-Quarter-Results.html}).

\begin{figure}[!ht]
	\centerline{\includegraphics[width=30em]{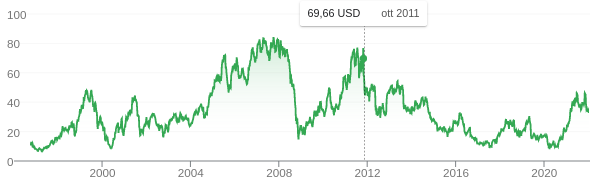}}
	\caption{ANF all trend.}
	\label{fig:marketCap}
\end{figure}
ANF was born at the end of September 1996, and from April to October 2011 it came close to the All Time High (ATH) at various times, meeting punctually a resistance. This peculiarity, together with the others previously mentioned, has led us to observe this stock with attention starting from the end of October 2011 (see fig. \ref{fig:marketCap}) when the downtrend period below \$70 begins, with the further intent to determine a potential opportune moment for a market entry by buying this asset.

\begin{figure}[!ht]
	\centerline{\includegraphics[width=45em]{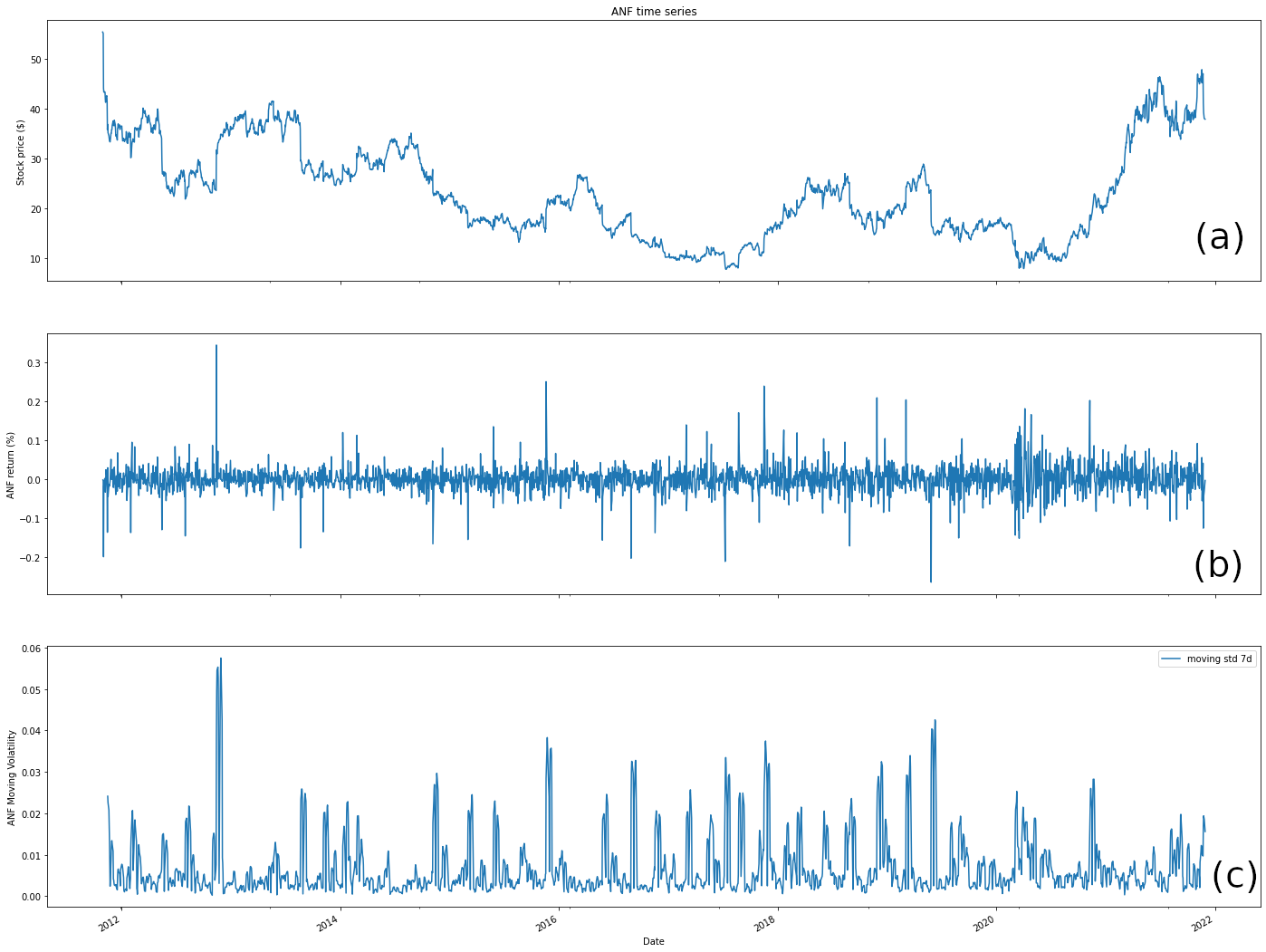}}
	\caption{ANF price (a), return (b), and volatility (c) from October 30th 2011 to November 30th 2021.}
	\label{fig:dataset}
\end{figure}

It is certainly an asset with a sometimes controversial trend and consequently well profitable if rightly analyzed, especially in the particular period of 2020/2021 due to the global pandemic (see fig. \ref{fig:dataset}(a)) and because it lacks the leverage effect (\cite{BlackBESS1976}) that explains the negative correlation between equity returns and return volatility (see fig. \ref{fig:dataset}(b) and \ref{fig:dataset}(c)) where most of the measures of the volatility of the asset are negatively correlated with its return.

Furthermore, in this long period analysed, ANF is not the classic always profitable stock (e.g. Tesla, Apple, Microsoft, or Bitcoin) to which trivially apply a passive Buy and Hold strategy\footnote{https://www.investopedia.com/terms/b/buyandhold.asp} for gains.

The dataset consists of the time series of the price of the above mentioned stocks, over the time period from October 16, 2011 to November 20, 2021, for a total period of 2537 open market days. The time series of price observations can be downloaded from Yahoo Finance\footnote{https://it.finance.yahoo.com/quote/ANF?p=ANF\&.tsrc} and then split into train and test sets, or directly from our github repository\footnote{https://github.com/IvanLetteri/StockTradingSystem-MLP-Regressor}.

\begin{figure}[!ht]
	\centerline{\includegraphics[width=35em]{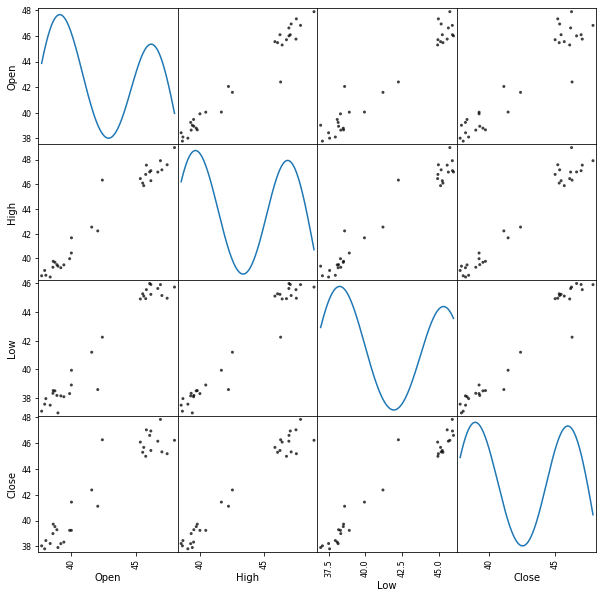}}
	\caption{OHLC scatter matrix of Validation set from 16th October to 20th November 2021).}
	\label{fig:scatter_valset}
\end{figure}
In particular, only the 30 days of the market open was used for the validation of the values composed by the features characterized by the OHLC prices aforementioned. In fig. \ref{fig:scatter_valset}, we can see the scatter matrix, which indicates that there are few positive correlations between them, even looking at the distribution of each individual variable along the diagonal represented with the Kernel Density Estimation (KDE) (see \cite{chen2017tutorial}), which denotes a smoother distribution surface across the plots. Intuitively, KDE has the effect of smoothing out each data point into a smooth bump. The shape of curve is determined by the kernel function $K(x)$ which sums overall the bumps to obtain the density estimator. The regions with many observations, KDE yield a large value because there are many bumps around. To the other side, for regions with few observations, the density value from summing over the bumps is low, because only few bumps contribute to the density estimate.

\subsubsection{Stationarity Test}
A time series is considered stationary when statistical properties such as mean, variance, and covariance are constant over time. For making predictions, the stationarity is a preferred characteristic, to the other side non-stationarity cannot exploit valuable time-dependent models, and variance can be misspecified by them.

To test whether our dataset is composed of a stationary time series, we used the stationary unit root as a statistical test. Given the time series $y_t=a*y_{t-1}+\epsilon_t$, where $y_t$ is the value at instant $t$ and $\epsilon_t$ is the error term. We evaluate, for all observations $y_t=a^p*y_{t-p}+\sum \epsilon_{t-i}*a^i$, if the value of $a$ is 1 (unity), then the predictions will equal $y_{t-p}$ and the sum of all errors from $t-p$ to $t$, meaning that the variance increases over time.

To verify the unit root stationary tests, we used the Augmented Dickey Fuller (ADF) test by analyzing the \textit{test statistic}, \textit{p-value}, and \textit{critical value} at 1\%, 5\%, and 10\% confidence intervals.

\begin{table}[!ht]
\centering
\caption{ADF test stationarity with AIC optimization.}
\label{tab:adf-test}
\begin{tabular}{l|c|c|c|c|ccc|}
\cline{2-8}
        & \textbf{Test Statistic} & \textbf{p-value} & \textbf{Lags} & \textbf{Observations} & \multicolumn{3}{c|}{\textbf{Critical Value}}                                                  \\ \cline{6-8} 
                                   &                                          &                                   &                                &                                        & \multicolumn{1}{c|}{\textbf{1\%}} & \multicolumn{1}{c|}{\textbf{5\%}} & \textbf{10\%}         \\ \hline
\multicolumn{1}{|l|}{\textit{ANF}} & -2.302                                   & 0.171                             & 5                              & 2529                                   & \multicolumn{1}{c|}{-3.432}       & \multicolumn{1}{c|}{-2.863}       & -2.567                \\ \hline
\end{tabular}
\end{table}
Tab. \ref{tab:adf-test} shows the number of lags considered, automatically selected based on the Akaike Information Criterion (AIC) (see \cite{Akaike1974H}) on ANF stock prices. The p-value result above the threshold (such as 5\% or 1\%) suggests rejecting the null hypothesis, so the time series turns out not to be stationary.

\section{Methodology}
In this paper, ARIMA model as a classical method and ANN as a deep learning model are chosen for ANF stocks price prediction. In this section, we analyze the models, comparing their results and selecting the best model based on the performance measures.

\subsection{Classic Method}
\label{sect:arima}
A statistical model is autoregressive (AR) if it predicts future values based on past values. By the term \textit{lag window}, we denote the set of previous observations (\textit{lags}) used to estimate the value at time $t$. For example, an autoregressive model of order $p$ ($AR(p)$), uses previous observations $p$ to predict the current value, so its lag window contains $y_{t-1}, \ldots, y_{t-p}$.

The Auto Correlation Function (ACF) shows how each observation is correlated $\phi_p = corr(y_t,y_{t-p})$ with its previous one, by measuring the linear relationship between observations at time $t$ and observations at the previous time $t-p$.

$$y_{t}=\phi_{1}y_{t-1}+ \dots + \phi_{p}y_{t-p}$$

\begin{figure}[!ht]
	\centerline{\includegraphics[width=40em]{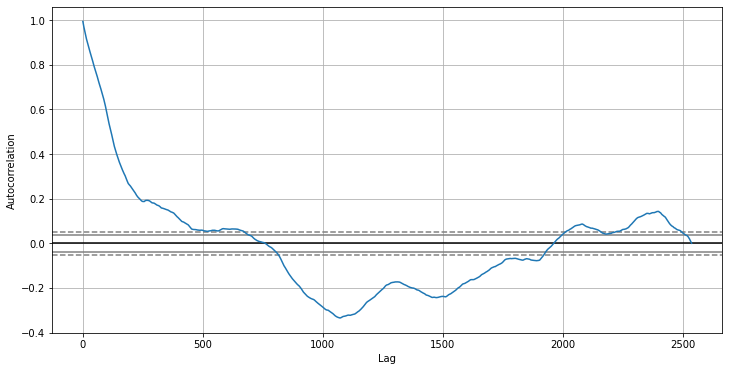}}
	\caption{ANF Values Correlation.}
	\label{fig:lags}
\end{figure}

\begin{figure}[!ht]
	\centerline{\includegraphics[width=40em]{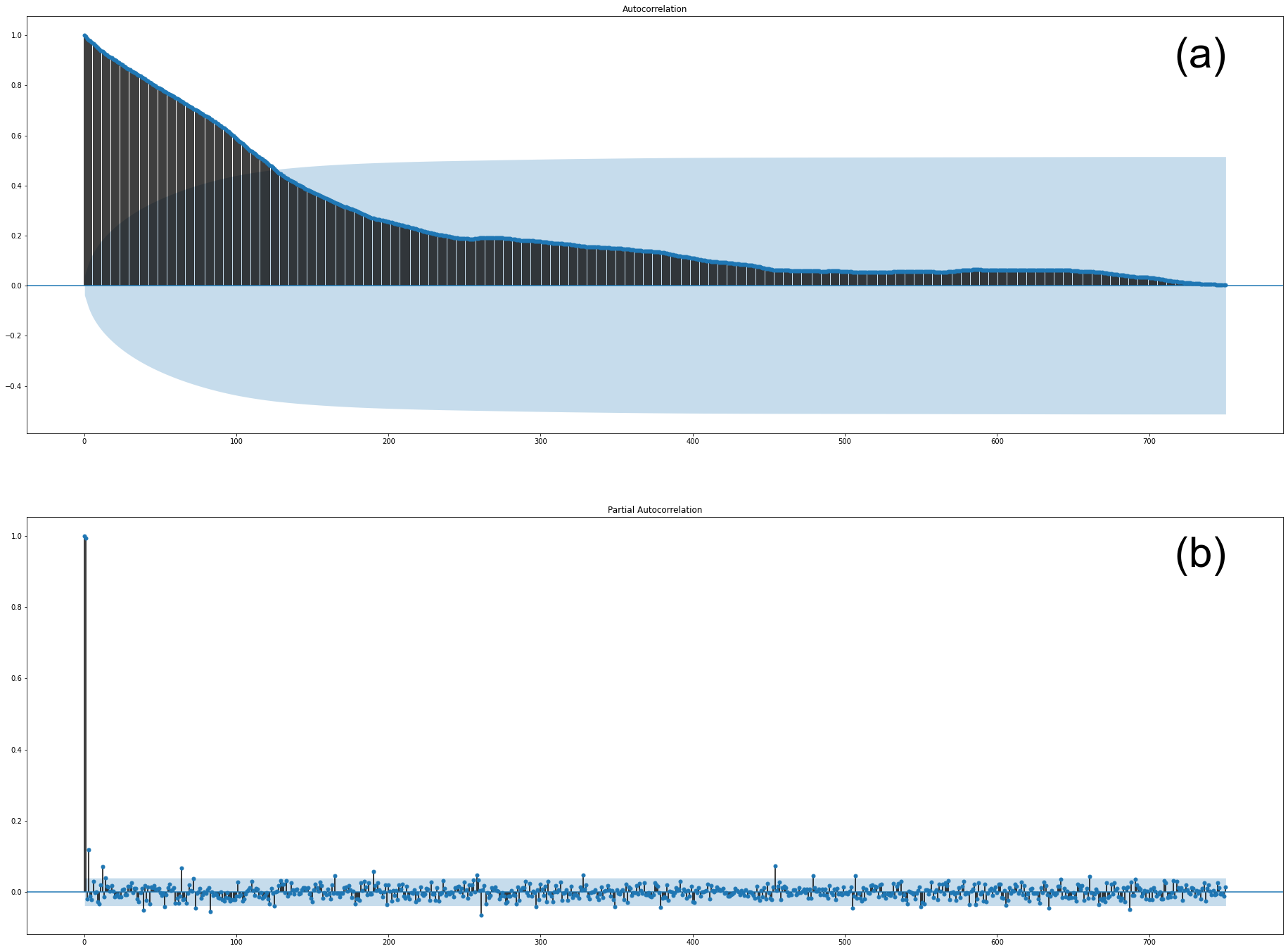}}
	\caption{ACF and Partial ACF of ANF price.}
	\label{fig:tsaANF}
\end{figure}

In our case, we can see from the ACF plot in Fig. \ref{fig:lags}, the plot of lag values on the $x$-axis and how they increase, and at the same time the correlation between price and lagged price on the $y$-axis deteriorates to $lag=750$. This means that the values of the historical series are strongly correlated with those of the lagged series for only an initial period, then the correlation decreases faster and faster (see fig. \ref{fig:tsaANF}(a)).

Although ACF shows that there is no high correlation between $y_{t-p}$ and $y_{t}$, it can happen that this correlation depends on the fact that both delays can be correlated with some intermediate delay and their correlation depends on that. To avoid this, we made further considerations by also evaluating the PACF that computes the correlation between $y_{t-p}$ and $y_{t}$ net of the intermediate variables (i.e., excluding correlations with them by $y_{t-p}$ and $y_{t}$).

Thus, if we transform the time series by filtering out the influence of observations at times $t-1 \ldots t-(p-1)$, leaving only the association between $y_t$ and $y_{t-p}$, then the ACF becomes the \textit{Partial Auto Correlation Function} (PACF), which denotes by $\alpha(p)$ a very small correlation (see fig. \ref{fig:tsaANF}(b)).

Since there is no specific trend in these plots, it can be assumed that the model was selected correctly and can be used for prediction.

Statistical models combined with machine learning have become ubiquitous in modern life. 

The well-known linear regression (LR) model has various forms, such as the Autoregressive (AR) model, the moving average (MA) model, the autoregressive moving average (ARMA) model, and its evolution, the autoregressive integrated moving average (ARIMA) model (\cite{Ham94}).

As a benchmark from a statistical point of view, we used the Auto-ARIMA model by \textit{pmdarima} \footnote{https://alkaline-ml.com/pmdarima/setup.html} library.
In the basic ARIMA model, we need to provide the $p$, $d$, and $q$ values using statistical techniques by performing the difference to eliminate the non-stationarity and plotting ACF and PACF graphs. In Auto-ARIMA, the model itself automatically discovers the optimal order for an ARIMA model by conducting differentiation tests (i.e., Kwiatkowski-Phillips-Schmidt-Shin, Augmented Dickey-Fuller, or Phillips-Perron) to determine the order of differentiation, $d$, and then fitting the models within defined ranges of beginning $p$, maximum $p$, beginning $q$, maximum $q$. If the seasonal option is enabled, Auto-ARIMA also attempts to identify the optimal hyper-parameters $P$ and $Q$ after conducting the Canova-Hansen (\cite{Eiji2002Kurozumi}) to determine the optimal order of seasonal differentiation, $D$.

Tab. \ref{tab:aic_arima} shows that the best model is ARIMA(0,1,0) by performing the stepwise search to minimize AIC (\cite{Stoica2004Selen}). Auto-ARIMA apply only one order of differenciation chosing the simple ARIMA called \textit{Random Walking} (see \cite{danyliv2019random}) with $p=0$, $d=1$, and $q=0$.

\begin{figure}[!ht]
	\centerline{\includegraphics[width=50em]{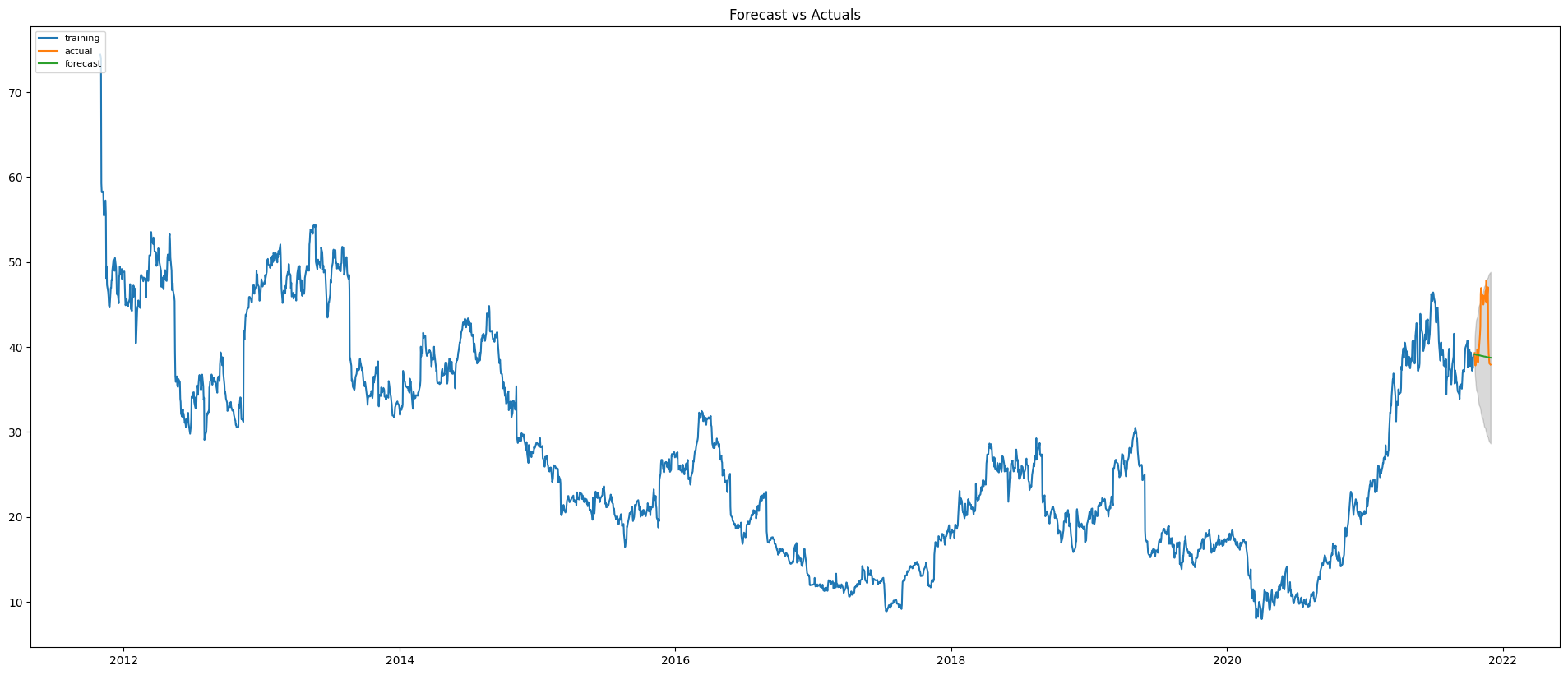}}
	\caption{ARIMA forecast of the last 30 days of ANF stock with p=0, d=1, q=0.}
	\label{fig:arima_forecat}
\end{figure}

\begin{table}[!ht]
\centering
\caption{ARIMA (p,d,q) models performing stepwise search to minimize Akaike Information Criterion.}
\label{tab:aic_arima}
\begin{tabular}{ccl}
\cline{1-2}
\textbf{ARIMA} & \textbf{AIC} &  \\ \cline{1-2}
(2,1,2)        & 6831.648     &  \\ \cline{1-2}
(0,1,0)        & 6830.378     &  \\ \cline{1-2}
(1,1,0)        & 6829.872     &  \\ \cline{1-2}
(0,1,1)        & 6829.776     &  \\ \cline{1-2}
\textbf{(0,1,0)}        & \textbf{6828.926}     &  \\ \cline{1-2}
(1,1,1)        & 6830.470     &  \\ \cline{1-2}
\end{tabular}
\end{table}
In fig. \ref{fig:arima_forecat} is shown the prediction made with the Auto ARIMA model to predict the ANF trend of the $n=30$ days following the training date. The training period falls in the time interval of 2507 days of market opening from October 30, 2011 to October 16, 2021.

\begin{figure}[!ht]
	\centerline{\includegraphics[width=47em]{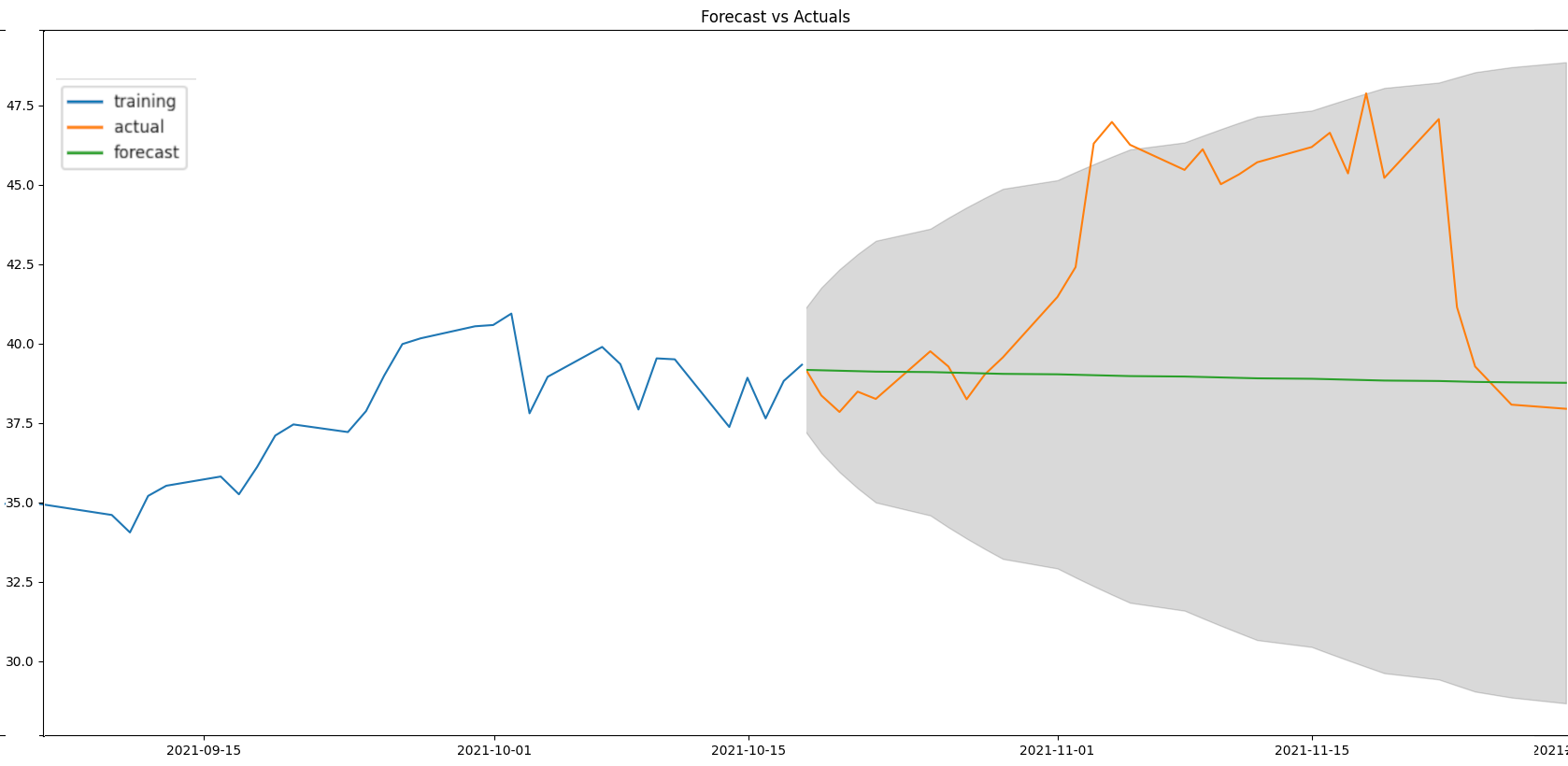}}
	\caption{Detail of ARIMA forecast of the last 30 days of ANF stock price.}
	\label{fig:detail_ARIMA}
\end{figure}

Respect to the DNN used in the next section \ref{sec:results}, ARIMA predictions are completely out of measure, as shown Tab. \ref{tab:error_arima}, relatively to the closing price. It not intercept any type of trend in the interval considered, as specified in fig. \ref{fig:detail_ARIMA}, and with detail in Tab. \ref{tab:values_ARIMA}). 

For the sake of brevity we omit the graphs of Low, High, and Open prices which are also far from the correct values. For this reason the next step was to turn to more refined models of machine learning, such as Deep Neural Networks (DNN).

\begin{table}[!ht]
\centering
\caption{ARIMA forecast of the last 30 days of ANF stock price trend.}
\label{tab:values_ARIMA}
\begin{tabular}{l|c|c|c|c|}
\cline{2-5}
\textbf{ARIMA}                        & \textbf{mean} & \textbf{std} & \multicolumn{1}{r|}{\textbf{min}} & \textbf{max} \\ \hline
\multicolumn{1}{|l|}{\textit{(p=0, d=1, q=0)}} & 38.94         & 0.123        & 38.73                             & 39.14        \\ \hline
\end{tabular}
\end{table}

\begin{table}[!ht]
\centering
\caption{Error metrics of ARIMA on ANF stock price prediction.}
\label{tab:error_arima}
\begin{tabular}{lccccc}
\hline
\textbf{ARIMA}  & \textbf{MSE} & \textbf{RMSE} & \textbf{MAE} & \textbf{MAPE} & \textbf{EVS}\\ \hline
\textit{(p=0, d=1, q=0)}         & 25.43    & 5.04      & 3.90         & 0.09         & -0.03\\ \hline
\end{tabular}
\end{table}

\subsection{Deep Learning Method}
Deep neural networks are a special type of artificial neural network characterized by an architecture consisting of a greater number of neurons and hidden layers than a conventional neural network can be. For each added hidden layer, in principle, the network will have a greater ability to extract high-level features.

DNNs have proven to be very efficient at solving nonlinear problems, including real-world problems. This is in contrast to many traditional techniques for time series forecasting, such as ARIMA, which assume that the series are generated by linear processes and consequently may be inappropriate for most real-world problems that are overwhelmingly nonlinear. While artificial neural networks (ANNs), as a soft computing technique, are the most accurate and widely used as prediction models in many areas, including social, engineering, economic, business, financial, foreign exchange and stock market problems.

An important work comparing the action prediction performance of ANN models versus ARIMA models was conducted by \cite{yao1999neural}, they showed that ANN model performed better than conventional ARIMA models. Equally decisive work was carried out by \cite{hansen1999time} they compared the prediction capabilities of ANNs and ARIMA focusing on time series prediction, they showed also that ANNs outperformed ARIMA in predicting the direction of stock movement as the latter was able to detect patterns hidden in the data used. 

From empirical results obtained, it was shown that ANN model is superior to ARIMA model and to these we add our work. Much other literature has demonstrated the prevalent use of ANNs as an effective tool for stock price prediction including \cite{GIORDANO20073871}. This makes ANNs a promising technique or potential hybrid for predicting time series movement. 

In this section, we maintain the same objective of forecasting OHLC values of the $n=30$ days following the training date, as already done with the ARIMA statistical model (see sect. \ref{sect:arima}). In this case, however, observations are grouped in batches of 5 days (batch-size $bs$), corresponding to the working days in an open market week.

When building a neural network suitable for financial applications, the main consideration is to find a compromise between generalisation and convergence. Of particular importance is to try not to have too many nodes in the hidden layer, as this could lead the DNN to learn only by not acquiring the ability to generalise.

The prediction method that determines the number of input neurons to DNNs is based on the \textit{lag windowing} criterion. This is a technique that consists in windowing the autocorrelation coefficients before estimating the linear prediction coefficients (LPC).

\begin{figure}[!ht]
	\centerline{\includegraphics[width=20em]{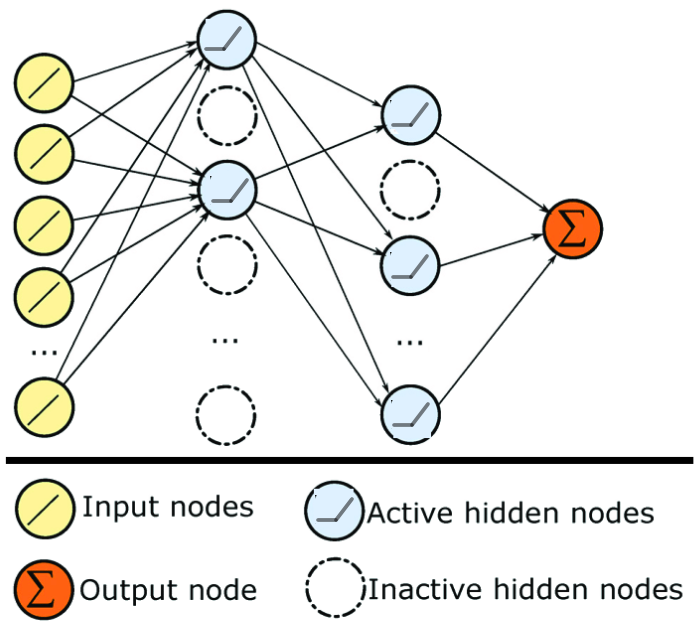}}
	\caption{Feed-forward DNN architecture.}
	\label{fig:dnn}
\end{figure}

For the generation of the feature target, we therefore used a lag window ($t=5$ days) identical to the optimal one found with the Auto ARIMA statistical model to maintain consistency for comparison. The size of the number of input neurons will be 5 for each of the four DNNs used on OHLC prices, and all with identical architecture. Actually, a very important part of this work, has been the identification of the best network geometry that allows the lowest error when applied on all 4 DNNs, which has been obtained by making a python module that generates different network geometries in combination with the sklearn GridSearchCV\footnote{https://scikit-learn.org/stable/modules/generated/sklearn.model\_selection.GridSearchCV.html}, for the setting of the hyper-parameters, propagated on all the DNNs. 
Specifically, following the input layer, we added 2 hidden layers composed of $n*t$ neurons in the first inner layer, reducing to $n*bs$ neurons in the second inner layer as in \cite{LetteriPG18} and in \cite{LetteriPG19}. Each neural network produces predictions of the price of the two considered stocks over the next $n$ days (see fig. \ref{fig:dnn}) via a single neuron in the output layer.

In Figure \ref{fig:dnn} is shown, an example of the network architecture used to which we applied a dropout of 0.2\% on each of the two internal layers to reduce overfitting (\cite{Geoffrey2012Hinton}). In addition, to introduce non-linearity between layers, we used ReLU as the activation function for overall neurons, which turns out to be better than a tanh or sigmoid (\cite{Krizhevsky2012Alex}), despite the fact that the depth of the network consists of only a few internal layers.

We applied an extension of stochastic gradient descent (SGD) called Adaptive moment (Adam) as an optimization algorithm to minimize the Loss function, in combination with the Back Propagation (BP) automatic differentiation algorithm to compute gradients of DNN weights, with respect to a training dataset.

The loss function applied is the Mean Absolute Error (MAE), also called \textit{L1 Loss}, since it is considered more robust to outliers, and useful in our case, since the price fluctuation of the asset ANF has some outliers due to its high volatility (see fig. \ref{fig:dataset}(c)).

Formally speaking, the MFA calculates the average of the sum of the absolute differences between the actual and predicted values as follows: $L1(y,\hat{y})=|y-\hat{y}|$, where $y$ represents the true value and $\hat{y}$ the predicted value. MAE checks the size of the errors in a set of predicted values, without worrying about their positive or negative direction. If the absolute values of the errors are not used, then the negative values may cancel out the positive values. 

All DNNs trained can be downloaded from our repository\footnote{https://www.ivanletteri.it/dnnModels/}.

\subsection{Multi-layer Perceptron Regressor Training}
\label{subsect:pts}
The strategies of a trading system tell the investor when to buy or sell shares in such a way that the sequence of these operations is profitable.

The trading system we propose makes forecasts of OHLC share prices for the \$30 open market days following the training date ending on October 16, 2021. The system's trading rules follow the following logic: when the expected closing price is higher than the current opening price, it is time to buy, and vice versa, to sell all the shares (when they are in the portfolio) at the moment when the expected closing price is lower. We have developed similar criteria considering the forecasts of the last month with the trained DNNs by defining a trading plan with a set of entry and exit rules.

We developed similar criteria by considering the forecasts of the last month with the trained DNNs defining a trading plan with a set of rules \textit{entry} and \textit{exit}.

\section{DNN Results}
\label{sec:results}
The evaluation is conducted through two components. The first one forecasts the OHLC prices of the asset, the second one calculates the performance of the trading system.

\subsection{Price Action Forecasting}
\begin{figure}[!ht]
	\centerline{\includegraphics[width=30em]{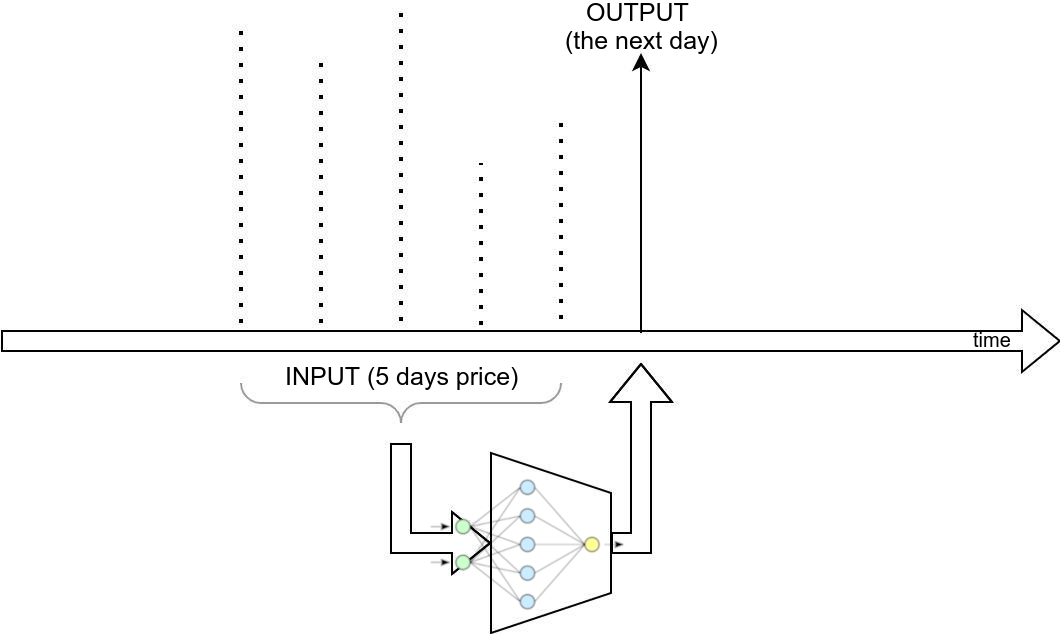}}
	\caption{DNN Input/Output.}
	\label{fig:nn_IO}
\end{figure}

The prediction procedure, of the 30 days of open market, is done by forecasting one day at a time starting from the 5 days known previous days, as shown in fig. \ref{fig:nn_IO}. The single neuron makes a single step forward, the network asks itself what the stock will do tomorrow, and this it does overall the next 30 days ahead using each time the data of the previous day.

Of continuation we bring back the graphs on the Price Action, in the 30 days successive the training, from which it is possible to compare the effective price regarding the previewed price from the DNNs.

\begin{figure}[!ht]
\centering
\begin{minipage}{.5\textwidth}
  \centering
  \includegraphics[width=1.\linewidth]{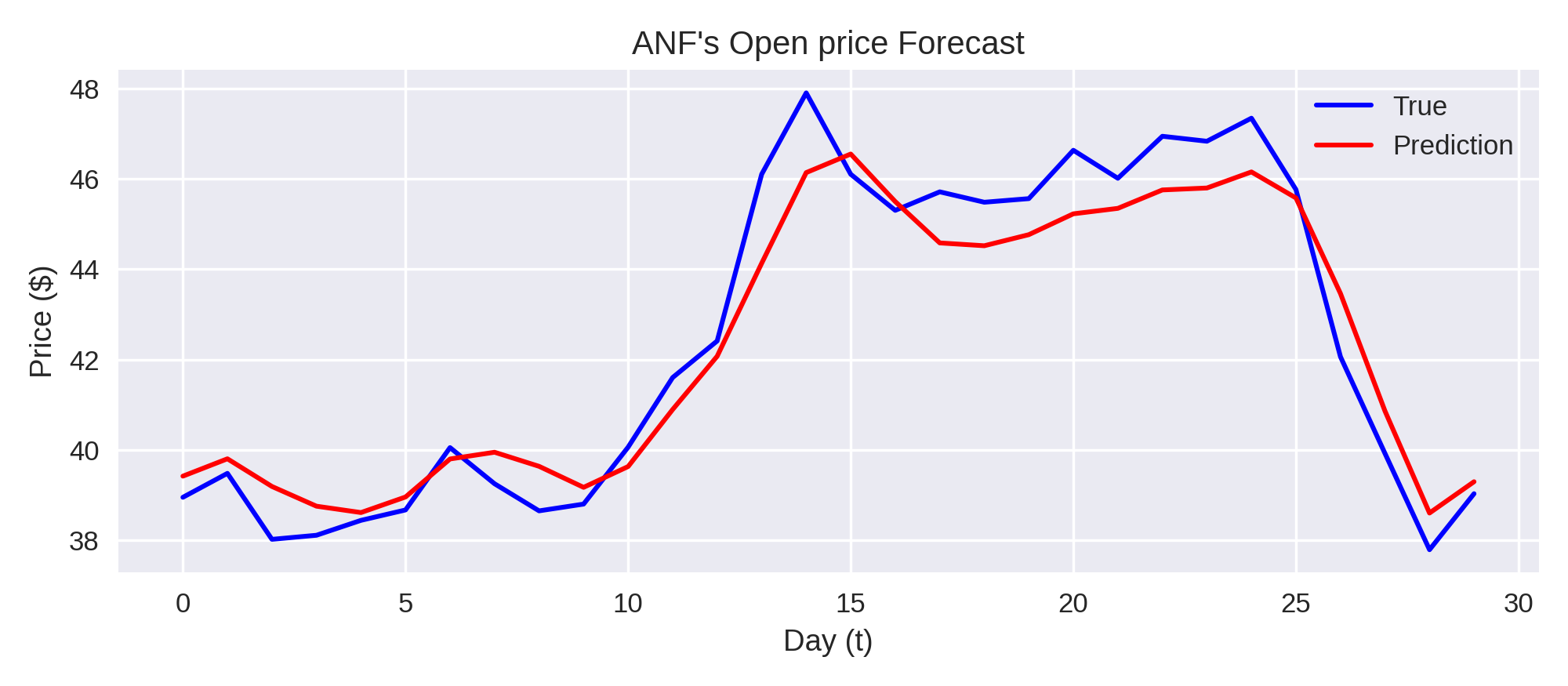}
  \label{fig:open}
\end{minipage}%
\begin{minipage}{.5\textwidth}
  \centering
  \includegraphics[width=1.\linewidth]{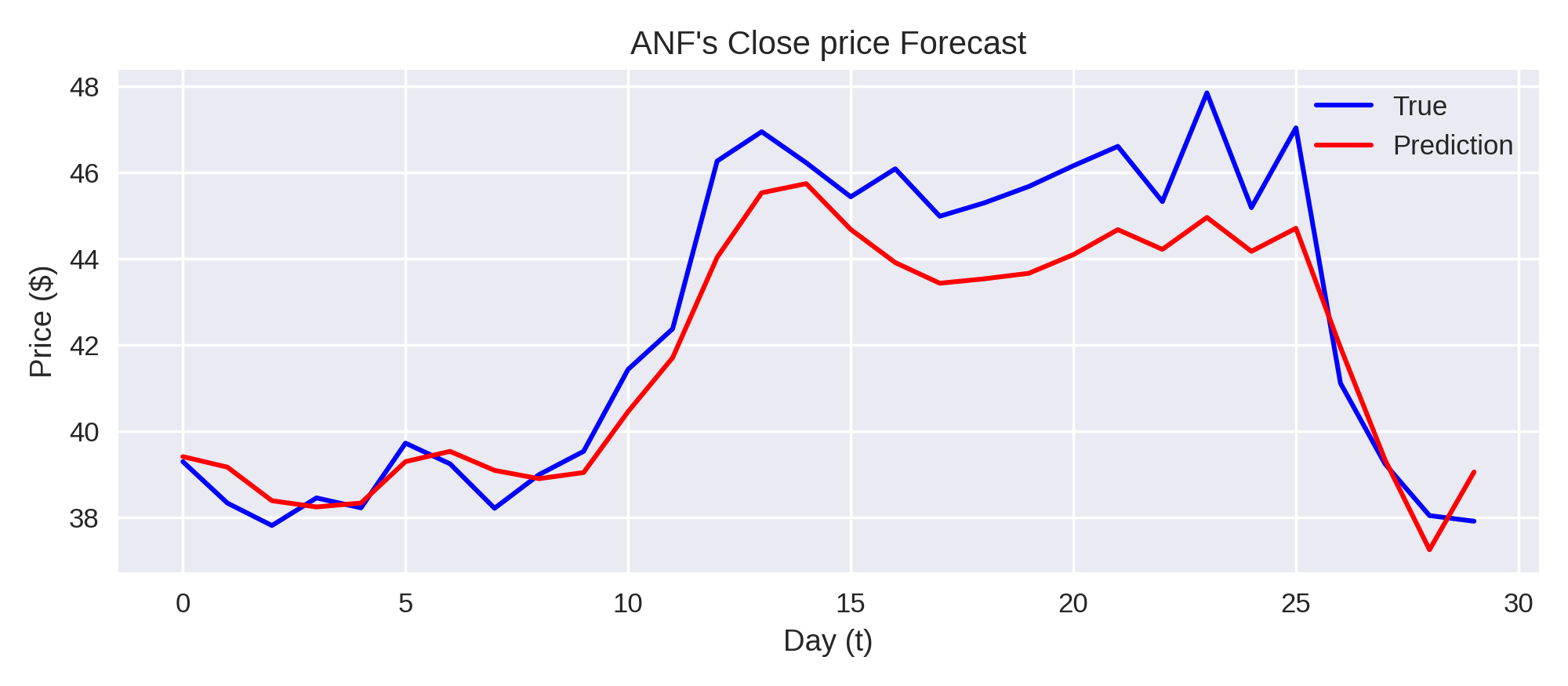}
  \label{fig:close}
\end{minipage}
\end{figure}

\begin{figure}[!ht]
\centering
\begin{minipage}{.5\textwidth}
  \centering
  \includegraphics[width=1.\linewidth]{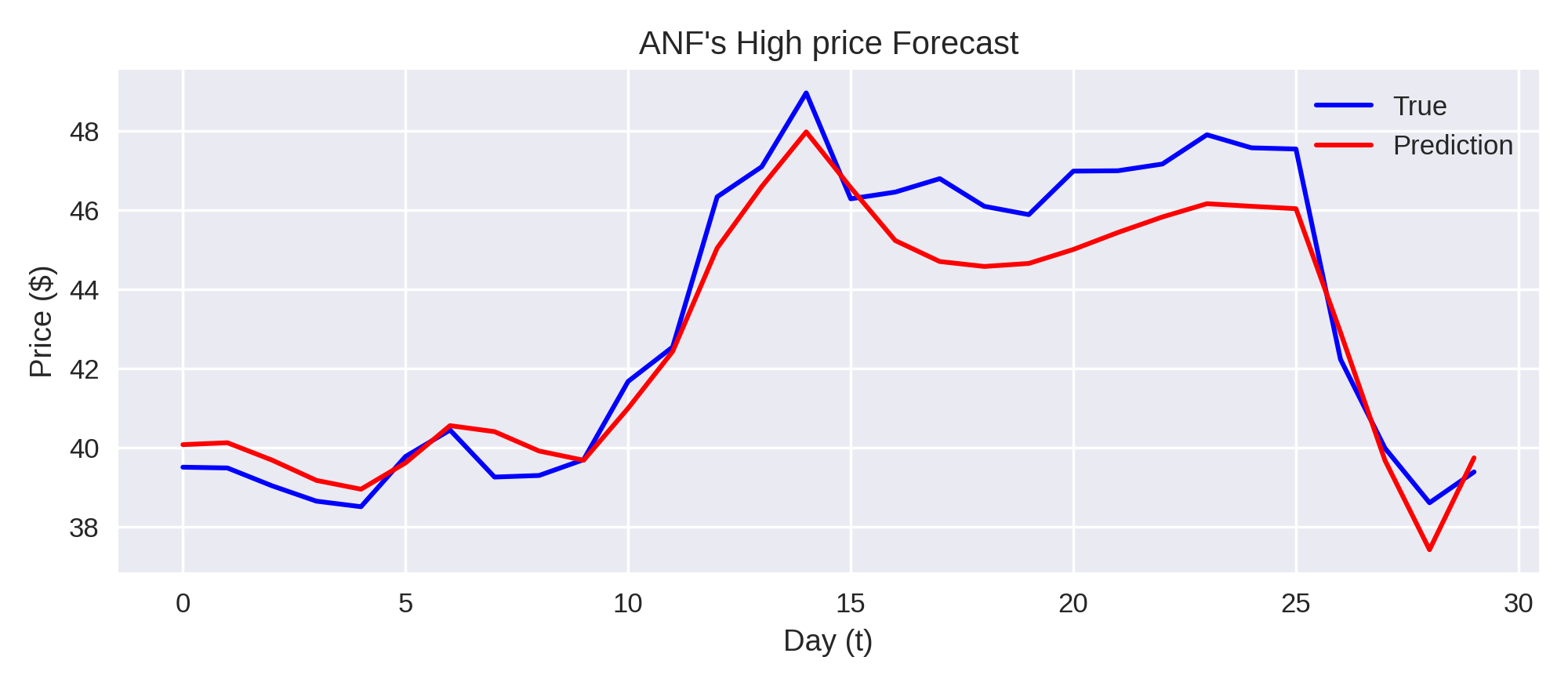}
  \label{fig:high}
\end{minipage}%
\begin{minipage}{.5\textwidth}
  \centering
  \includegraphics[width=1.\linewidth]{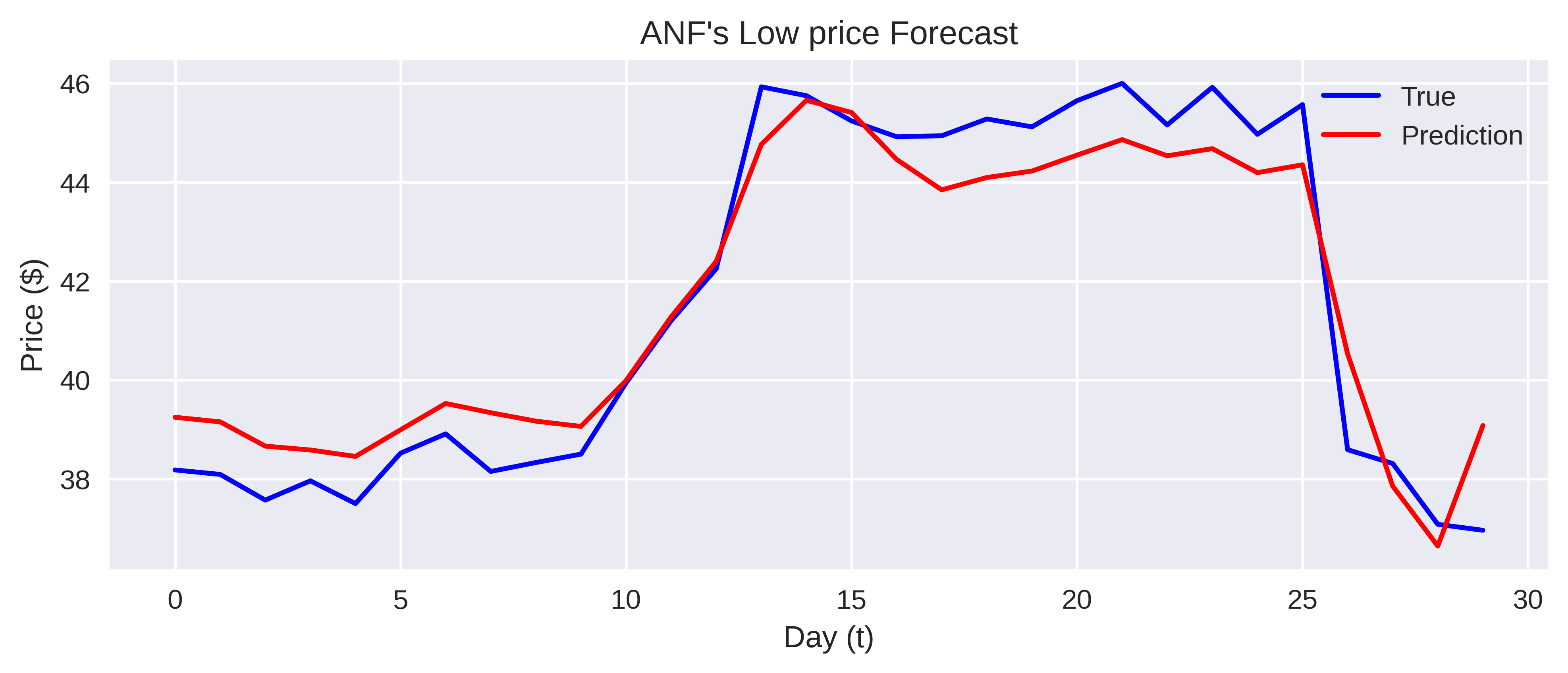}
  \label{fig:low}
\end{minipage}
\caption{DNNs prediction of OHLC prices on validation set.}
\end{figure}

\subsubsection{Error Metrics}
The evaluation metrics used for the predictive models are:

\begin{itemize}
    \item \textit{Mean Square Error} (MSE): is calculated as the mean of the squared forecast error values. The square root of the forecast error values dictates that they are positive and induces the effect of giving greater weight to larger errors. When the forecast errors are very large or outliers are squared, which has the effect of dragging the mean of the squared forecast errors outwards, this converges to a larger mean square error score. The MSE tends to give a worse score to models that make large forecast errors. The MSE is expressed as follows: 
    $$MSE(y,\hat{y})=\frac{1}{N}\sum_{t=1}^N(y_t-\hat{y_t})^2$$
    \item \textit{Root Mean Square Error} (RMSE): is the standard deviation of the residuals (prediction errors). Residuals are a measure of how far from the regression line data points are. RMSE measures of how spread out these residuals are as follow:
    $$RMSE(y,\hat{y})=\sqrt{\sum_{t=1}^N(y_t-\hat{y_t})^2}.$$
    \item \textit{Mean Absolute Error} (MAE): is a measure of the errors between pairs of observations expressing the same phenomenon, calculated as: $$MAE(y,\hat{y})=\frac{1}{N}\sum_{t=1}^N|(y_t-\hat{y_t})|$$
    \item \textit{Mean Absolute Percentage Error} (MAPE): measures this accuracy as a percentage and can be calculated as the average absolute percentage error for each time period minus the actual values divided by the actual values as follows:
    $$MAPE(y,\hat{y})=\frac{1}{N}\sum_{t=1}^N\big|\frac{y_t-\hat{y_t}}{y_t}\big|$$
    where: $n$ is the number of fitted points, $y$ is the actual value, $\hat{y}$ is the forecast value. $\sum$ is summation notation (the absolute value is summed for every forecasted point in time). MAPE is the most common measure used to forecast error, and works best if there are no extremes to the data (and no zeros).
    \item \textit{Explained Variance Score} (EVS): explains the dispersion of errors of a given dataset as follows: $$EVS(y,\hat{y})=1-\frac{Var(y-\hat{y})}{Var(y)}$$
    where $Var(y-\hat{y})$,  and $Var(y)$ is the variance of prediction errors and actual values respectively. Scores close to 1.0 are highly desired, indicating better squares of standard deviations of errors.
\end{itemize}
 
Predictive accuracy of all four neural network model evaluated as the coefficient determination i.e. MSE, RMSE, MAE, MAPE. All results reported are for the test set. The predictive estimation results are summarized in Table \ref{tab:error_comparison}.

\begin{table}[!ht]
\centering
\caption{Error metrics of ANF stock price prediction of validation set.}
\label{tab:error_comparison}
\begin{tabular}{lccccc}
\hline
\textbf{DNN}  & \textbf{MSE} & \textbf{RMSE} & \textbf{MAE} & \textbf{MAPE} & \textbf{EVS}\\ \hline
Open         & 0.82    & 0.91      & 0.77         & 0.02         & 0.94\\ \hline
High         & 1.14    & 0.08      & 0.90         & 0.02         & 0.93\\ \hline
Low          & 0.93    & 0.97      & 0.83         & 0.02         & 0.93\\ \hline
Close        & 1.75    & 1.32      & 1.07         & 0.02         & 0.91 \\ \hline
\end{tabular}
\end{table}

\section{Backtracking Trading System}
Given a trading strategy that defines entry and exit points, backtesting is the process of evaluating the performance of a trading strategy based on historical data by analysing price charts to find the gross return, recording all trades, etc...

\subsection{Profit and Risk Metrics}
In this section, we set out the metrics related to downside risk by estimating the potential loss in value of the stock.

\begin{itemize}
    \item \textit{Maximum drawdown (MDD)}\footnote{https://www.investopedia.com/terms/m/maximum-drawdown-mdd.asp} which measures the largest decline from the peak in the whole trading period to show the worst case as follow: $MDD=max_{\tau \in (0,t)}[max_{t \in (0,\tau)}\frac{n_t-n_{\tau}}{n_t}]$.
    \item \textit{Sharpe ratio (SR)}\footnote{https://www.investopedia.com/terms/s/sharperatio.asp} as the risk-adjusted profit measure, which refers to the return per unit of deviation as follow: $SR = \frac{\mathbb{E}[r]}{[r]}$.
    \item \textit{Sortino ratio (SoR)}\footnote{https://www.investopedia.com/terms/s/sortinoratio.asp} is a variant of risk-adjusted profit measure, which applies DD as risk measure: $SoR = \frac{\mathbb{E}[r]}{DD}$.
    \item \textit{Calmar ratio (CR)}\footnote{https://www.investopedia.com/terms/c/calmarratio.asp} is another variant of risk-adjusted profit measure, which applies MDD as risk measure: $CR = \frac{\mathbb{E}[r]}{MDD}$.
\end{itemize}
To check the goodness of trades, we mainly focused on the \textit{Total Returns} $R_{k}(t)$ for each ANF stocks $(k = 1, ...,p)$ in the time interval $(t= 1, ...,n)$ 
    \begin{equation}
        R_{k}(t) = \frac{Z_k(t+\Delta t) - Z_{k}(t)}{Z_{k}(t)}
    \end{equation}
and furthermore analysing the standardized returns $r_k = (R_k - \mu_k) / \sigma_k,$ with $(k = 1, ...,p)$, where $\sigma_k$ is the standard deviation of $R_k$, e $\mu_k$ denote the average overtime for the studied period.

\subsection{Triple EMA strategy for Entry and Exit rules}
For the selection of trading strategies we tested a set of 12 technical indicators, three for each of the following areas: trend, momentum, volatility, and volume indicators. The best of them all turned out to be the Triple EMA. The Triple EMA strtegy is a trend following system, based of the Exponential Moving Average (EMA) $EMA_{n_t} = EMA_{t-1}+(\frac{2}{n+1})[x_t^{(ohlc)} - EMA_{t-1}]$.

For the sake of brevity we omit the results of the other 11 indicators, obtained after the backtracking tests, reporting only the best (Triple EMA). The triple EMA (\cite{Tsantekidis2017Avraam}) is generally used to make short-term and medium-term intraday trading decisions to enter long or short trades, depending on bullish or bearish signals. Below is the formula for the triple exponential moving average:
$$Triple EMA_{n_t} = (3 * EMA_{n_t}) - (3 * EMA(EMA_{n_t})) + EMA(EMA(EMA_{n_t}))$$
Determined the Triple EMA based on the expected OHLC with the DNNs, we define the entry/exit rules as follows:
\begin{itemize}
    \item \textbf{Entry rule} \\
    if $(x_{n_t}^{(l)} < Triple EMA_{n_t}^{(l)})$ or $(x_{n_t}^{(h)} < Triple EMA_{n_t}^{(h)})$ and \\
    $(x_{n_t}^{(c)} < Triple EMA_{n_t}^{(c)})$ or $(x_{n_t}^{(o)} < Triple EMA_{n_t}^{(o)})$ is \textbf{True}.
    \item \textbf{Exit rule} \\
    if$(x_{n_t}^{(l)} > Triple EMA_{n_t}^{(l)})$ or $(x_{n_t}^{(h)} > Triple EMA_{n_t}^{(h)})$ and \\
    $(x_{n_t}^{(c)} > Triple EMA_{n_t}^{(c)})$ or $(x_{n_t}^{(o)} > Triple EMA_{n_t}^{(o)})$ is \textbf{True}.
\end{itemize}
In this paper, we have used a backtracking system to estimate: the total profit from a budget of \$100 reinvested in compounded mode by calculating the Expectancy Ratio (ER) as follows: $ER = WR * \mu_{win} - (1 - WR) * | \mu_{loss}|$, where $WR$ is the win rate that evaluates the proportion of the trading period with positive profit among all trading intervals. The $\mu_{win}$ is the average of profitable trades and $\mu_{loss}$ is the average of the break even\footnote{https://www.investopedia.com/terms/b/breakevenanalysis.asp}. All of these metrics, with details about Sharp, Sortino, and Calmar ratios, are shown in the tab. \ref{tab:results}.

\begin{figure}[!ht]
\centering
\begin{minipage}{.5\textwidth}
  \centering
  \includegraphics[width=1.\linewidth]{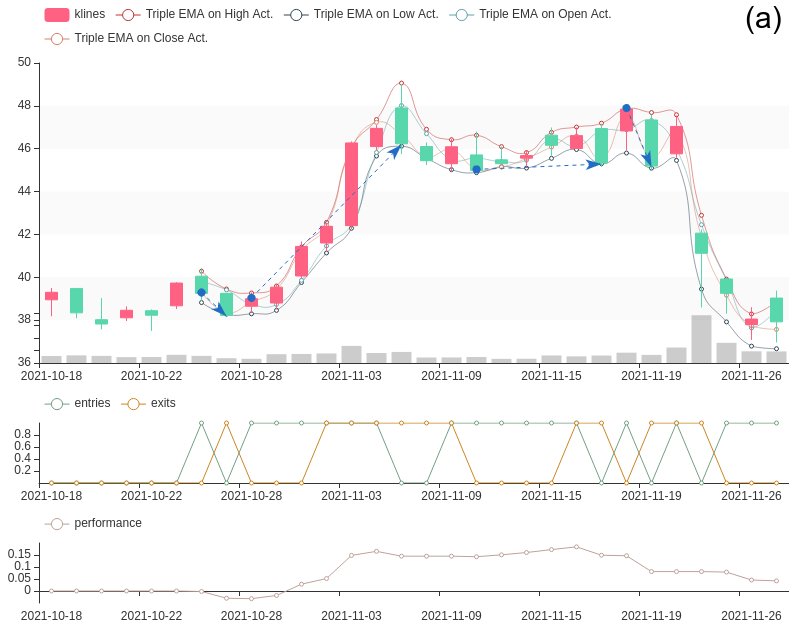}
  \label{fig:trades_actual}
\end{minipage}%
\begin{minipage}{.5\textwidth}
  \centering
  \includegraphics[width=1.\linewidth]{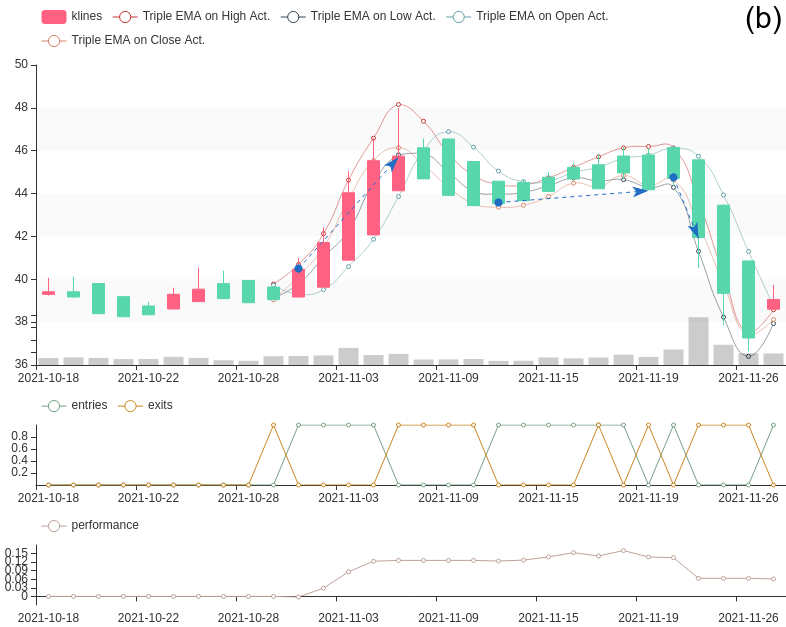}
  \label{fig:trades_predicted}
\end{minipage}
\caption{Performance results on real OHLC values on the left and predicted OHLC values to the right.}
\label{fig:trades}
\end{figure}

\begin{table}[!ht]
\centering
\caption{Performance of trading system using different entry rules.}
\label{tab:results}
\begin{tabular}{lcccccc}
\hline
\textbf{Validation set} & \textbf{\begin{tabular}[c]{@{}c@{}}Num. of\\ Trades\end{tabular}} & \textbf{\begin{tabular}[c]{@{}c@{}}Total\\ Return (\$)\end{tabular}} & \textbf{\begin{tabular}[c]{@{}c@{}}Expectancy\\ Ratio \end{tabular}} & \textbf{\begin{tabular}[c]{@{}c@{}}Sharpe\\ Ratio\end{tabular}} & \textbf{\begin{tabular}[c]{@{}c@{}}Sortino\\ Ratio\end{tabular}} & \textbf{\begin{tabular}[c]{@{}c@{}}Calmar\\ Ratio\end{tabular}} \\ \hline
actual        & 4         & 4.220           & 2.022                       & 1.280              & 2.220                         & 5.459              \\ \hline
predicted    & 3         & 6.126           & 2.112                       & 2.194              & 3.340                        & 12.403              \\ \hline
\end{tabular}
\end{table}
From the tab. \ref{tab:results}, we can see that the system, applying the Triple EMA to the OHLC price with real values, makes 4 trades (see fig. \ref{fig:trades}(a)), while using the Triple EMA on the OHLC prices predicted by the DNN the system chooses to reduce the number of trades to three. The latter is more effective, resulting in a higher total profit.

In detail, we have that on 3 trades (see fig. \ref{fig:trades}(b)) a Win Rate of 66.6667\%, a Best Trade 12.611\%, Worst Trade -6.5578 \%, and a Avg. Trade 2.37017\%. We also verified that the strategy wins on the classic Buy \& Hold quote, which has a negative return of -0.904755\%.

After determining the best trades, based on the Triple EMA, we applied them to the forecast obtained with the DNNs. Subsequently, these same trades were applied to the actual data and calculated the Cumulative Total Profit.

\begin{figure}[!ht]
	\centerline{\includegraphics[width=35em]{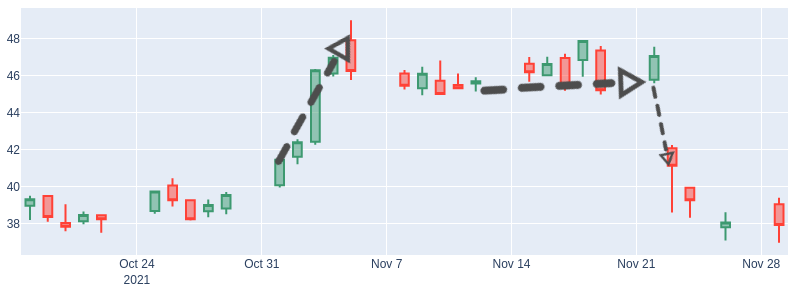}}
	\caption{Trades predicted by DNN mapped on real OHLC price of the validation set.}
	\label{fig:trades_pred}
\end{figure}
In fig. \ref{fig:trades_pred}, we have mapped the 3 trades onto the actual values of the validation set and reported in tab. \ref{tab:pred_trades} the dates of the trades to the market price in the period of the trades. In particular, but the sum of the profits and losses deriving from the spread of each trade allowed us to obtain a total return of 3.2\% in one month on an invested capital of \$100.

\begin{table}[!ht]
\centering
\caption{Details of trades.}
\begin{tabular}{l|c|c|c|}
\cline{2-4}
\textbf{Trades}         & \textbf{\begin{tabular}[c]{@{}c@{}}entry\\ {[} date | price(\$){]}\end{tabular}} & \textbf{\begin{tabular}[c]{@{}c@{}}exit\\ {[} date | price(\$){]}\end{tabular}} & \multicolumn{1}{r|}{\textbf{Profit (\$)}} \\ \hline
\multicolumn{1}{|l|}{1} & 1st Nov. | 40.46                                                                 & 5th Nov. | 45.74                                                                & 5.28                                      \\ \hline
\multicolumn{1}{|l|}{2} & 11th Nov. | 43.36                                                                & 19th Nov. | 44.13                                                               & 0.77                                      \\ \hline
\multicolumn{1}{|l|}{3} & 22th Nov. | 44.75                                                                & 23th Nov. | 41.90                                                               & -2.85                                     \\ \hline
\end{tabular}
\label{tab:pred_trades}
\end{table}

\newpage

\section{Conclusion}
In this paper, we propose a stock market trading system that uses four deep neural networks as part of its main components. After a forecasting activity with the regressive Auto ARIMA model, which proved inconclusive, we used DNNs with OHLC prices from historical data to forecast the trend of the ANF stock in the 30 days of open market, following the training date. These four ML models are incorporated into the trading system to make automated buy and sell decisions using a robot advisor.

For the sake of simplicity, we have omitted experiments conducted by reducing and increasing the number of days to predict, because they showed that predictions further into the future produced less accurate results, whereas 30 days is a good compromise and on average gives the trading system sufficient time to identify a trade.

The consequences of changing the trading rules were also studied by testing multiple technical indicators and combinations of them. The best one turned out to be the Triple EMA, which allowed us to make a profit of \$3.2 in one month by applying the trades obtained from crossing the 3-period Triple EMA with the OHLC prices predicted by the DNNs, starting from a budget of only \$100.

In the near future we plan to explore additional technical indicators and more exploratory data analysis following a proven process (see \cite{LetteriPVG20}). The new datasets will certainly require a balancing of buy, sell and hold trades. This balancing will take place according to the algorithm called \textit{Generative 1 Nearest Neighbours} (\cite{LetteriArxivG1No}) and targeted oversampling (\cite{LetteriCDP21}).

Furthermore, we plan to study other types of deep learning algorithms by combining RNN and CNN. We also plan to apply better trading rules that can use more advanced methods of assigning stop limits and take profits. XAI criteria inherent to feature selection as in \cite{LetteriArxivG1No, DyoubCLL20} will be established. 

Artificial Intelligence (AI) advisors in general, and AI buy \& sell advisors in particular, introduce ethical issues and concerns. There behaviours should be monitored for ethical violation (\cite{DyoubCLL21}, \cite{DyoubICLP21}). As another future work, we plan to consider the ethical aspects related to AI buy \& sell advisors. 

Given the enormous potential and thanks also to the extendability and robustness of this trading system, we will further test its reliability on other stock markets using other inputs to the dataset such as cryptocurrencies and defi-tokens, also varying the timeframes for day trading and scalping activities.

\newpage

\bibliographystyle{unsrtnat}
\bibliography{references}  

\begin{thebibliography}{27}
\providecommand{\natexlab}[1]{#1}
\providecommand{\url}[1]{\texttt{#1}}
\expandafter\ifx\csname urlstyle\endcsname\relax
  \providecommand{\doi}[1]{doi: #1}\else
  \providecommand{\doi}{doi: \begingroup \urlstyle{rm}\Url}\fi

\bibitem[Letteri et~al.(2020{\natexlab{a}})Letteri, Penna, Vita, and
  Grifa]{LetteriPVG20}
Ivan Letteri, Giuseppe~Della Penna, Luca~Di Vita, and Maria~Teresa Grifa.
\newblock Mta-kdd'19: {A} dataset for malware traffic detection.
\newblock 2597:\penalty0 153--165, 2020{\natexlab{a}}.
\newblock URL \url{http://ceur-ws.org/Vol-2597/paper-14.pdf}.

\bibitem[Letteri et~al.(2018)Letteri, Penna, and Gasperis]{LetteriPG18}
Ivan Letteri, Giuseppe~Della Penna, and Giovanni~De Gasperis.
\newblock Botnet detection in software defined networks by deep learning
  techniques.
\newblock 11161:\penalty0 49--62, 2018.
\newblock \doi{10.1007/978-3-030-01689-0\_4}.
\newblock URL \url{https://doi.org/10.1007/978-3-030-01689-0\_4}.

\bibitem[Letteri et~al.(2019{\natexlab{a}})Letteri, Penna, and
  Gasperis]{LetteriPG19}
Ivan Letteri, Giuseppe~Della Penna, and Giovanni~De Gasperis.
\newblock Security in the internet of things: botnet detection in
  software-defined networks by deep learning techniques.
\newblock \emph{Int. J. High Perform. Comput. Netw.}, 15\penalty0
  (3/4):\penalty0 170--182, 2019{\natexlab{a}}.
\newblock \doi{10.1504/IJHPCN.2019.106095}.
\newblock URL \url{https://doi.org/10.1504/IJHPCN.2019.106095}.

\bibitem[Letteri et~al.(2019{\natexlab{b}})Letteri, Penna, and
  Caianiello]{LetteriPC19}
Ivan Letteri, Giuseppe~Della Penna, and Pasquale Caianiello.
\newblock Feature selection strategies for {HTTP} botnet traffic detection.
\newblock pages 202--210, 2019{\natexlab{b}}.
\newblock \doi{10.1109/EuroSPW.2019.00029}.
\newblock URL \url{https://doi.org/10.1109/EuroSPW.2019.00029}.

\bibitem[Yetis et~al.(2014)Yetis, Kaplan, and Jamshidi]{Yetis2014StockMP}
Yunus Yetis, Halid Kaplan, and Mo~M. Jamshidi.
\newblock Stock market prediction by using artificial neural network.
\newblock \emph{2014 World Automation Congress (WAC)}, pages 718--722, 2014.

\bibitem[Soniya et~al.(2015)Soniya, Paul, and Singh]{Soniya2015ARO}
Soniya, Sandeep Paul, and Lotika Singh.
\newblock A review on advances in deep learning.
\newblock \emph{2015 IEEE Workshop on Computational Intelligence: Theories,
  Applications and Future Directions (WCI)}, pages 1--6, 2015.

\bibitem[Day and Lee(2016)]{Day2016DeepLF}
Min-Yuh Day and Chia-Chou Lee.
\newblock Deep learning for financial sentiment analysis on finance news
  providers.
\newblock \emph{2016 IEEE/ACM International Conference on Advances in Social
  Networks Analysis and Mining (ASONAM)}, pages 1127--1134, 2016.

\bibitem[Letteri et~al.(2020{\natexlab{b}})Letteri, Cecco, Dyoub, and
  Penna]{LetteriDSopt2020}
Ivan Letteri, Antonio~Di Cecco, Abeer Dyoub, and Giuseppe~Della Penna.
\newblock A novel resampling technique for imbalanced dataset optimization.
\newblock \emph{CoRR}, abs/2012.15231, 2020{\natexlab{b}}.
\newblock URL \url{https://arxiv.org/abs/2012.15231}.

\bibitem[Wang et~al.(2021)Wang, Huang, and Wang]{wang2021forecasting}
Huiwen Wang, Wenyang Huang, and Shanshan Wang.
\newblock Forecasting open-high-low-close data contained in candlestick chart,
  2021.

\bibitem[Black(1976)]{BlackBESS1976}
F.~Black.
\newblock Studies of stock price volatility changes.
\newblock 3002:\penalty0 177--181, 1976.

\bibitem[Chen(2017)]{chen2017tutorial}
Yen-Chi Chen.
\newblock A tutorial on kernel density estimation and recent advances, 2017.

\bibitem[Akaike(1974)]{Akaike1974H}
H.~Akaike.
\newblock A new look at the statistical model identification.
\newblock \emph{IEEE Transactions on Automatic Control}, 19\penalty0
  (6):\penalty0 716--723, 1974.
\newblock \doi{10.1109/TAC.1974.1100705}.

\bibitem[Marquez(1995)]{Ham94}
Jamie Marquez.
\newblock Time series analysis : James d. hamilton, 1994, (princeton university
  press, princeton, nj).
\newblock \emph{International Journal of Forecasting}, 11\penalty0
  (3):\penalty0 494--495, September 1995.

\bibitem[Kurozumi(2002)]{Eiji2002Kurozumi}
Eiji Kurozumi.
\newblock The limiting properties of the canova and hansen test under local
  alternatives.
\newblock \emph{Econometric Theory}, 18\penalty0 (5):\penalty0 1197--1220,
  2002.
\newblock ISSN 02664666, 14694360.
\newblock URL \url{http://www.jstor.org/stable/3533370}.

\bibitem[Stoica and Selen(2004)]{Stoica2004Selen}
P.~Stoica and Y.~Selen.
\newblock Model-order selection: a review of information criterion rules.
\newblock \emph{IEEE Signal Processing Magazine}, 21\penalty0 (4):\penalty0
  36--47, 2004.
\newblock \doi{10.1109/MSP.2004.1311138}.

\bibitem[Danyliv et~al.(2019)Danyliv, Bland, and Argenson]{danyliv2019random}
Oleh Danyliv, Bruce Bland, and Alexandre Argenson.
\newblock Random walk model from the point of view of algorithmic trading,
  2019.

\bibitem[Yao et~al.(1999)Yao, Tan, and Poh]{yao1999neural}
Jingtao Yao, Chew~Lim Tan, and Hean-Lee Poh.
\newblock Neural networks for technical analysis: a study on klci.
\newblock \emph{International journal of theoretical and applied finance},
  2\penalty0 (02):\penalty0 221--241, 1999.

\bibitem[Hansen et~al.(1999)Hansen, McDonald, and Nelson]{hansen1999time}
James~V Hansen, James~B McDonald, and Ray~D Nelson.
\newblock Time series prediction with genetic-algorithm designed neural
  networks: An empirical comparison with modern statistical models.
\newblock \emph{Computational Intelligence}, 15\penalty0 (3):\penalty0
  171--184, 1999.

\bibitem[Giordano et~al.(2007)Giordano, {La Rocca}, and
  Perna]{GIORDANO20073871}
Francesco Giordano, Michele {La Rocca}, and Cira Perna.
\newblock Forecasting nonlinear time series with neural network sieve
  bootstrap.
\newblock \emph{Computational Statistics \& Data Analysis}, 51\penalty0
  (8):\penalty0 3871--3884, 2007.
\newblock ISSN 0167-9473.
\newblock \doi{https://doi.org/10.1016/j.csda.2006.03.003}.
\newblock URL
  \url{https://www.sciencedirect.com/science/article/pii/S0167947306000740}.

\bibitem[Hinton et~al.(2012)Hinton, Srivastava, Krizhevsky, Sutskever, and
  Salakhutdinov]{Geoffrey2012Hinton}
Geoffrey~E. Hinton, Nitish Srivastava, Alex Krizhevsky, Ilya Sutskever, and
  Ruslan Salakhutdinov.
\newblock Improving neural networks by preventing co-adaptation of feature
  detectors.
\newblock \emph{CoRR}, abs/1207.0580, 2012.
\newblock URL \url{http://arxiv.org/abs/1207.0580}.

\bibitem[Krizhevsky et~al.(2012)Krizhevsky, Sutskever, and
  Hinton]{Krizhevsky2012Alex}
Alex Krizhevsky, Ilya Sutskever, and Geoffrey~E. Hinton.
\newblock Imagenet classification with deep convolutional neural networks.
\newblock In \emph{Proceedings of the 25th International Conference on Neural
  Information Processing Systems - Volume 1}, NIPS'12, page 1097–1105, Red
  Hook, NY, USA, 2012. Curran Associates Inc.

\bibitem[Tsantekidis et~al.(2017)Tsantekidis, Passalis, Tefas, Kanniainen,
  Gabbouj, and Iosifidis]{Tsantekidis2017Avraam}
Avraam Tsantekidis, Nikolaos Passalis, Anastasios Tefas, Juho Kanniainen,
  Moncef Gabbouj, and Alexandros Iosifidis.
\newblock Forecasting stock prices from the limit order book using
  convolutional neural networks.
\newblock In \emph{2017 IEEE 19th Conference on Business Informatics (CBI)},
  volume~01, pages 7--12, 2017.
\newblock \doi{10.1109/CBI.2017.23}.

\bibitem[Letteri et~al.(2020{\natexlab{c}})Letteri, Cecco, and
  Penna]{LetteriArxivG1No}
Ivan Letteri, Antonio~Di Cecco, and Giuseppe~Della Penna.
\newblock Dataset optimization strategies for malware traffic detection.
\newblock \emph{CoRR}, abs/2009.11347, 2020{\natexlab{c}}.
\newblock URL \url{https://arxiv.org/abs/2009.11347}.

\bibitem[Letteri et~al.(2021)Letteri, Cecco, Dyoub, and Penna]{LetteriCDP21}
Ivan Letteri, Antonio~Di Cecco, Abeer Dyoub, and Giuseppe~Della Penna.
\newblock Imbalanced dataset optimization with new resampling techniques.
\newblock In Kohei Arai, editor, \emph{Intelligent Systems and Applications -
  Proceedings of the 2021 Intelligent Systems Conference, IntelliSys 2021,
  Amsterdam, The Netherlands, 2-3 September, 2021, Volume 2}, volume 295 of
  \emph{Lecture Notes in Networks and Systems}, pages 199--215. Springer, 2021.
\newblock \doi{10.1007/978-3-030-82196-8\_15}.
\newblock URL \url{https://doi.org/10.1007/978-3-030-82196-8\_15}.

\bibitem[Dyoub et~al.(2020)Dyoub, Costantini, Lisi, and Letteri]{DyoubCLL20}
Abeer Dyoub, Stefania Costantini, Francesca~Alessandra Lisi, and Ivan Letteri.
\newblock Logic-based machine learning for transparent ethical agents.
\newblock In Francesco Calimeri, Simona Perri, and Ester Zumpano, editors,
  \emph{Proceedings of the 35th Italian Conference on Computational Logic -
  {CILC} 2020, Rende, Italy, October 13-15, 2020}, volume 2710 of \emph{{CEUR}
  Workshop Proceedings}, pages 169--183. CEUR-WS.org, 2020.

\bibitem[Dyoub et~al.(2021{\natexlab{a}})Dyoub, Costantini, Lisi, and
  Letteri]{DyoubCLL21}
Abeer Dyoub, Stefania Costantini, Francesca~A. Lisi, and Ivan Letteri.
\newblock Ethical monitoring and evaluation of dialogues with a mas.
\newblock 3002:\penalty0 158--172, 2021{\natexlab{a}}.

\bibitem[Dyoub et~al.(2021{\natexlab{b}})Dyoub, Costantini, Letteri, and
  Lisi]{DyoubICLP21}
Abeer Dyoub, Stefania Costantini, Ivan Letteri, and Francesca~A. Lisi.
\newblock A logic-based multi-agent system for ethical monitoring and
  evaluation of dialogues.
\newblock In Andrea Formisano, Yanhong~Annie Liu, Bart Bogaerts, Alex Brik,
  Ver{\'{o}}nica Dahl, Carmine Dodaro, Paul Fodor, Gian~Luca Pozzato, Joost
  Vennekens, and Neng{-}Fa Zhou, editors, \emph{Proceedings 37th International
  Conference on Logic Programming (Technical Communications), {ICLP} Technical
  Communications 2021, Porto (virtual event), 20-27th September 2021}, volume
  345 of \emph{{EPTCS}}, pages 182--188, 2021{\natexlab{b}}.
\newblock \doi{10.4204/EPTCS.345.32}.
\newblock URL \url{https://doi.org/10.4204/EPTCS.345.32}.

\end{thebibliography}




\end{document}